\documentclass[prl,twocolumn]{revtex4}
\usepackage{graphicx}

\begin{document}

\title{Zero-Point cooling and low heating of trapped $^{111}$Cd$^{+}$ ions}
\author{L. Deslauriers, P. C. Haljan, P. J. Lee, K-A. Brickman, B. B. Blinov, M. J. Madsen and C. Monroe}
\affiliation{FOCUS Center, Department of Physics, University of Michigan}

\date{\today}

\begin{abstract}
We report on ground state laser cooling of single $^{111}$Cd$^+$ ions confined in radio-frequency (Paul) traps. Heating rates of trapped ion motion are measured for two different trapping geometries and electrode materials, where no effort was made to shield the electrodes from the atomic Cd source. The low measured heating rates suggest that trapped $^{111}$Cd$^{+}$ ions may be well-suited for experiments involving quantum control of atomic motion, including applications in quantum information science. 
\end{abstract}
\maketitle

Trapped ions are widely accepted as one of the most promising architectures for realizing a quantum information processor \cite{ciraczoller:1995,wineland:1998}. Particular internal states of an ion can behave as an ideal quantum bit, and entanglement between multiple ions can proceed through Coulomb interaction mediated by appropriate laser fields. Many entangling schemes require the ions to be initialized to near the $n=0$ ground state of motion \cite{diedrich:1989,monroe1:1995,turchette1:1998,wineland:1998,schmidt-kaler:2003,leibfried:2003,gill:2004}, and motional decoherence caused by anomalous heating of trapped ion motion \cite{turchette:2000} can be a limiting factor in the fidelity of quantum logic gates. Effective zero-point cooling of trapped ion motion and suppression of motional heating are thus crucial to many applications of trapped ions to quantum information science.

We report the Raman-sideband cooling of $^{111}$Cd$^{+}$ ions to the ground state of motion and heating of the ion's motion from fluctuating background electric fields. These measurements are conducted in two different trap structures with different geometries, trap strengths, sizes, and electrode materials.

Recent work with trapped $^{9}$Be$^{+}$ ions suggested that the cleanliness and the condition of electrode surfaces - in particular, the plating of the electrodes from the atomic source used to load ions - can play an important role in trapped-ion heating \cite{rowe:2001,turchette:2000}. Related work with trapped $^{137}$Ba$^{+}$ ions demonstrated a gradual degradation of trap stability as the electrodes became coated with barium \cite{devoe:2002}. In the $^{111}$Cd$^{+}$ system, we observe heating rates comparable to the lowest heating rates observed in the $^{9}$Be$^{+}$ system, without shielding the electrodes from the atomic source. These measurements have taken place over six months of continuous operation and revealed no noticeable change in behavior. This suggests that cadmium-coated electrodes might not adversely affect the heating or stability of trapped ions.

\begin{figure}[hptb]
\centering
\includegraphics[angle=0,width=8.6cm,keepaspectratio]{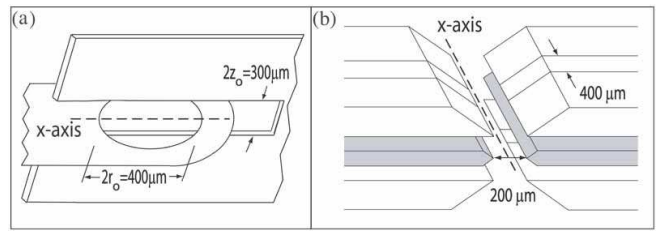}
\caption{Schematic diagrams for the electrodes of the two ion traps used in this experiment. (a) Asymmetric quadrupole trap with closest electrode distance $d$=150 $\mu$m (fork electrode). Both electrodes are constructed from 125 $\mu$m thick molybdenum sheets. (b) Three-layer linear trap with closest electrode distance $d$=100 $\mu$m (middle rf electrode). The rf layer (gray) is 125 $\mu$m thick, while the top and bottom segmented layers are 250 $\mu$m thick; the gold-plated alumina layers are separated by alumina spacers (not shown) with a thickness of 125 $\mu$m. The outer segments of the top and the bottom layers are separated by 400~$\mu$m. The evaporated gold coating on each electrode is approximately 0.3 $\mu$m thick. The electrodes in both traps are not shielded from the atomic Cd sources.  } 
\label{fig:trapsfinal}
\end{figure}

We confine single $^{111}$Cd$^+$ ions in two different radio-frequency (rf) Paul traps \cite{dehmelt:1967}. The first  trap is an asymmetric quadrupole trap similar to that described in refs.~\cite{jefferts:1995,blinov:2002}. It consists of a ring electrode with a radius of $r_0~\cong$~200~$\mu$m and a fork electrode with a gap of $2z_0~\cong$~300~$\mu$m, as shown in Fig.~\ref{fig:trapsfinal}(a). We apply a potential $V(t)=V_{0}\cos(\Omega_{T}t)$ to the fork electrode and a static potential $U_0$ to the ring electrode. With $V_{0}\approx400$~V, $\Omega_{T}/2\pi\approx50$ MHz and $U_0=30~V$, the pseudopotential trap frequencies \cite{dehmelt:1967} are $(\omega_x,\omega_y,\omega_z)/2\pi \approx (5.8,8.9,9.7)$~MHz. For $U_0=6$~V, the trap frequencies are $(\omega_x,\omega_y,\omega_z)/2\pi \approx (4.8,7.4,11.3)$~MHz. The second trap is a three-layer linear trap  consisting of gold-coated alumina substrates vertically stacked with 125 $\mu$m alumina spacers between them, as shown in Fig. \ref{fig:trapsfinal}(b). The outer electrodes are axially segmented into three sections with appropriate static potentials applied to each for axial confinement and for compensation of background static electric fields. The axial trap frequency ranges from about 400~kHz to 4~MHz as the potential difference between the outer and the inner segments varies from 5~V to 275~V. For transverse confinement, an rf potential with amplitude $V_{0}\approx400$~V with respect to the outer electrode layers is applied to the middle electrode layer, resulting in transverse trap frequencies of about 8~MHz. Cadmium ions are loaded by directing ultraviolet laser radiation ($10^{6}$ mW/cm$^{2}$ at 214.5 nm) onto the trap electrodes while also evaporating cadmium from nearby Cd or CdO sources. We speculate that the ultraviolet radiation either dislodges Cd$^+$ ions directly from the surface or ejects photoelectrons which ionize Cd atoms in the trap region.

\begin{figure}[hptb]
\centering
\includegraphics[angle=0,width=8.6cm,keepaspectratio]{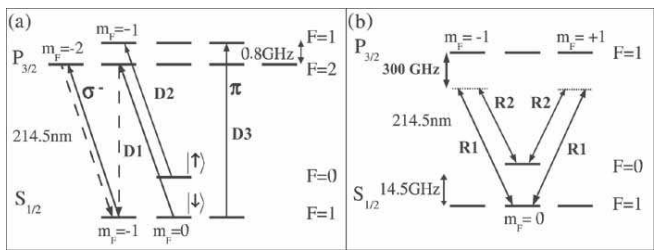}
\caption{(a) Relevant $^{111}$Cd$^+$ energy levels (not to scale). The ground state hyperfine levels used in the experiment are $\left|\downarrow\right\rangle$ = $\left|1,0 \right\rangle$  and $\left|\uparrow\right\rangle$ = $\left|0,0 \right\rangle $ (expressed in the $\left|F,M_{F}\right\rangle$ basis). (b) Both Raman beams R1 and R2 are linearly polarized and are red detuned 300~GHz from the $S_{1/2} \rightarrow P_{3/2}$ transition. } 
\label{fig:heating_fig1}
\end{figure}

The~relevant~level~structure~of~$^{111}$Cd$^{+}$~(nuclear spin I=1/2)~is~shown~in~Fig.~\ref{fig:heating_fig1}.~The two ground state hyperfine levels used in the experiment are the $5s^{2}S_{1/2}\left|F=0,m_F=0\right\rangle$$\equiv\left|\uparrow\right\rangle$ and $\left|F=1,m_F=0\right\rangle$$\equiv\left|\downarrow\right\rangle$ states, forming an effective qubit separated in frequency by $\omega_{HF}/2\pi~=~14.53$~GHz.~Three near-resonant laser beams at wavelengths near 214.5 nm drive transitions between the $^{2}$$S_{1/2}$ ground states and $^{2}P_{3/2}$ electronic excited states, as depicted in Fig.~\ref{fig:heating_fig1}(a). Beam D1 drives the cycling transition $^{2}$$S_{1/2}\left|F=1,m_F=-1\right\rangle$~$\rightarrow$~$^{2}$$P_{3/2}\left|F=2,m_F=-2\right\rangle$ and provides Doppler laser cooling when accompanied by beam D2, which is tuned to the~$\left|\uparrow\right\rangle$~$\rightarrow$~$^{2}$$P_{3/2}\left|F=1,m_F=-1\right\rangle$ transition and prevents accumulation in the $\left|\uparrow\right\rangle$ state.~Beam D3 prepares the $^{111}$Cd$^{+}$ ion in the $\left|\uparrow\right\rangle$ state through optical pumping. Beam D1 alone allows the detection of the qubit state through standard quantum jump techniques \cite{blattzoller:1988}: when the ion is prepared in state $\left|\uparrow\right\rangle$, beam D1 is far from resonance and results in very little ion fluorescence; when the ion is prepared in state $\left|\downarrow\right\rangle$, it is optically pumped to the $S_{1/2}\left|F=1,m_F=-1\right\rangle$ state with high probability and thereby results in high fluorescence.  We observe a qubit detection efficiency of 99.7 $\%$ for the $^{111}$Cd$^{+}$ qubit.

Laser beams R1 and R2 shown in Fig.~\ref{fig:heating_fig1}(b) are used to drive resolved stimulated Raman transitions between  $\left|\uparrow\right\rangle$$\left|n\right\rangle$ and $\left|\downarrow\right\rangle$$\left|n'\right\rangle$ for 1-D sideband cooling \cite{monroe1:1995}. Here, $\left|n\right\rangle$ denotes the ion's quantum state of harmonic motion along the $x$-axis with an energy E$_{n}$=$\hbar\omega_{x}(n+1/2)$ where $n=0,1,2,3\ldots$. In the quadrupole trap, the $x$-axis is the weak axis in the plane of the ring, and in the linear trap, the $x$-axis is along the axial dimension of the linear trap, as shown in Fig.~\ref{fig:trapsfinal}. The Raman beams are detuned 300~GHz below the $S_{1/2}$ to $P_{3/2}$ transition as shown in Fig. \ref{fig:heating_fig1}(b). We derive the Raman beams using an electro-optic phase modulator, from which optical sideband pairs having a beatnote near $\omega_{HF}$ interfere to give an effective stimulated Raman process \cite{lee:2003}; fine tuning of the beatnote frequency is accomplished with acousto-optic modulators. The wave-vector difference $\vec{\delta k}$ of the Raman beams is set parallel to the $x$-axis so that the Raman transitions are sensitive only to the ion's motion along this direction, with a corresponding Lamb-Dicke parameter of $\eta_{x} = 0.28\nu^{-1/2}$, where $\nu=\omega_{x}/2\pi$ is the trap frequency in MHz. When the beatnote is tuned to the qubit resonance frequency, the Raman ``carrier" transition $(n'=n)$ is driven, exhibiting a Rabi frequency $\Omega_{0}/2\pi \approx$ 100~kHz. When the Raman beatnote frequency is increased by $\delta=+\omega_{x}$ with respect to the carrier, the first upper sideband transition $(n'=n-1)$ is driven, and has Rabi frequency $\eta_{x}\Omega_{0}\sqrt{n}$. Likewise, by decreasing the Raman beatnote to $\delta=-\omega_{x}$ with respect to the carrier , the first lower sideband transition $(n'=n+1)$ is driven, with Rabi frequency $\eta_{x}\Omega_{0}\sqrt{n+1}$. (These expressions for Rabi frequency are valid in the Lamb-Dicke regime, when $\eta_{x}\sqrt{n+1} \ll 1$).

Ground state cooling is achieved in two steps: First, the ion is Doppler cooled with laser beams D1 and D2. The Doppler laser beams have components along all principal axes of the trap, hence provide cooling in all three dimensions. This first cooling stage brings the population to a thermal distribution near the Doppler limit with an average vibrational number $\bar{n}_{D} \approx  \gamma_{o}/2\omega_{x}$ \cite{wineland:1979}, where $\gamma_{o}/2\pi \cong$~47 MHz is the $P_{3/2}$ excited state linewidth. A second cooling stage is performed using sideband Raman cooling cycles between the two qubit states $\left|\downarrow\right\rangle$ and $\left|\uparrow\right\rangle$ \cite{monroe1:1995}. In each cycle of Raman cooling, the ion is driven on the first upper Raman sideband, transferring population from $\left|\uparrow\right\rangle\left|n\right\rangle$ to $\left|\downarrow\right\rangle\left|n-1\right\rangle$. The qubit is then optically pumped from $\left|\downarrow\right\rangle$ to the state $\left|\uparrow\right\rangle$ using beam D3, which on average does not appreciably change the motional state of the ion. These two steps are alternated many times, eventually resulting in the cooling to near the $\left|n=0\right\rangle$ ground state of motion.

\begin{figure}[hptb]
\centering
\includegraphics[angle=0,width=8cm,keepaspectratio]{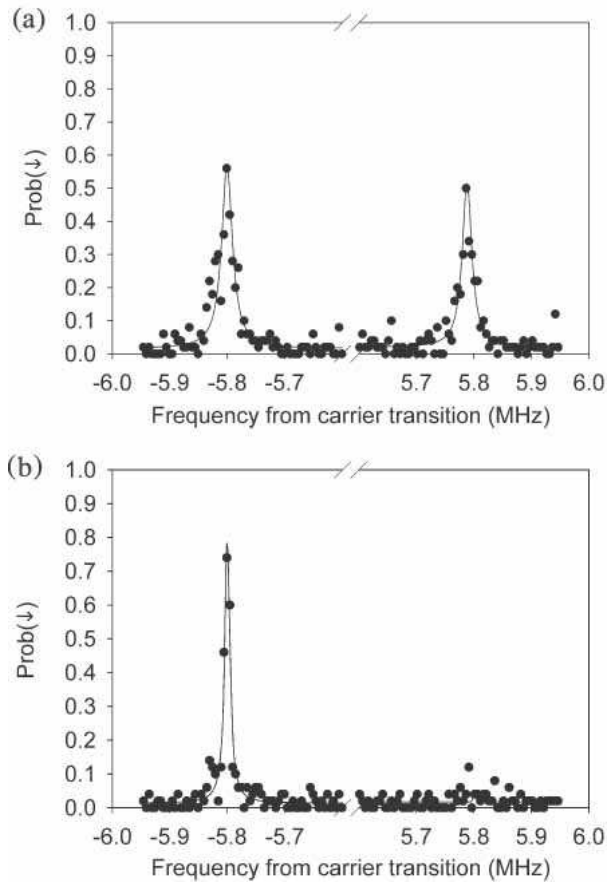}
\caption{Raman spectra for an ion in the quadrupole trap with a secular frequency of $\omega_{x}/2\pi$=5.8 MHz. Both plots show probability for population transfer to the ``bright state" P($\left|\downarrow\right\rangle~ =~\left|F=1,M_{F}=0\right\rangle$) vs $\delta$ or frequency of the beatnote from a carrier transition. Both lower(left) and upper(right) sidebands are displayed following (a) Doppler cooling to $\bar{n}$~$\simeq$~5(3), and (b) subsequent Raman cooling to $\bar{n}$~$\simeq$~0.03(2) where the upper sideband vanishes. The strength of the transitions are $\Omega_{0}/2\pi$~=~100 kHz and $\Omega_{0,1}/2\pi$ = 10 kHz. The Raman probe pulse is exposed for 80 $\mu$sec. The lines are a fit to the data points}
\label{fig:ramanspectrum}
\end{figure}

The average vibrational number $\bar{n}$ is extracted through the measured asymmetry in the first upper and first lower sideband strengths; for a thermal distribution of vibrational levels, the ratio of the sideband strengths is $\bar{n}/(1+\bar{n})$ \cite{monroe1:1995}. Raman spectra for the first upper and lower sidebands are shown for Doppler cooling (Fig. \ref{fig:ramanspectrum}(a)) and subsequent Raman cooling (Fig. \ref{fig:ramanspectrum}(b)) in the quadrupole trap with $\omega_{x}/2\pi$ = 5.8 MHz ($\eta_{x} \simeq$ 0.12). The change in the sideband asymmetries indicates cooling from approximately $\bar{n}\simeq$ 5(3) to $\bar{n}\simeq$ 0.03(2), corresponding to a probability P$_{0}$ $\simeq$ 97$\%$ of ground state occupation. No further cooling is observed when more than about 40 Raman cooling cycles are used, and the results are largely independent of the details of the Raman cooling schedule. For instance, a uniform setting for the Raman cooling sideband pulse works nearly as well as lengthening the Raman pulses appropriately as the ion is cooled \cite{monroe1:1995}. Similar results are observed in both the quadrupole and the linear traps for various frequencies between 1.3 MHz to 5.8 MHz. In the linear trap with $\omega_{x}/2\pi$ = 2.69 MHz ($\eta_{x} \simeq$ 0.17), we reach a probability P$_{0}\simeq$ 83$\%$ of ground state occupation, requiring no more than 90 Raman cooling cycles.

The efficacy of first-sideband Raman cooling relies upon Doppler cooling to near the Lamb-Dicke limit, i.e. $\eta_{x}^{2}\bar{n}_{D} \lesssim 1$. Assuming the Doppler limit is reached in the first cooling stage, the required trap strength for effective Doppler/Raman first-sideband cooling to the ground state can be roughly written as $\omega_{x} \gtrsim \sqrt{\gamma_{o} \omega_{R}/2}$, where $\omega_{R} = \hbar(\delta k)^{2}/2m$ is the recoil frequency associated with the transfer of optical momentum of the Raman transition. Interestingly, this trap frequency threshold for effective Doppler/Raman cooling is roughly  $\omega_{x}/2\pi \sim 1$~MHz for many ion species (including $^{9}$Be$^{+}$, $^{40}$Ca$^{+}$, and $^{111}$Cd$^{+}$).

\begin{figure}[hptb]
\centering
\includegraphics[angle=0,width=8.6cm,keepaspectratio]{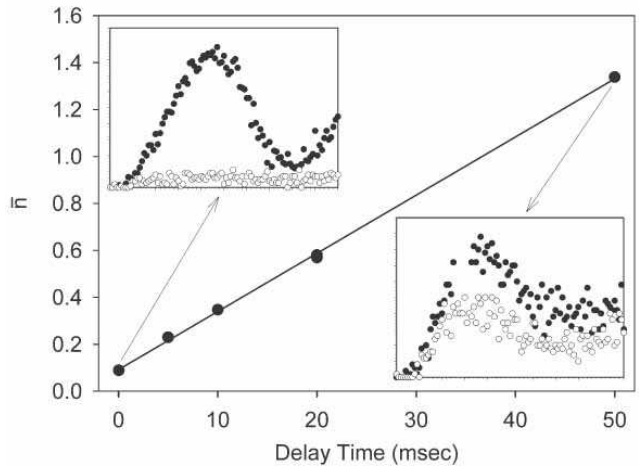}
\caption{An example of heating data taken in the quadrupole trap, with trap frequency $\omega_{x}/2\pi$=5.8 MHz. Mean motional quanta $\bar{n}$ is plotted vs time delay. The insets show sideband Rabi oscillations from which $\bar{n}$ is inferred; the black points represent the lower sideband and the open points represent the upper sideband. The solid line is a linear fit to the data from which a heating rate of $\dot{\bar{n}}$ $\approx$ 0.0248(3) quanta/msec is obtained.  }
\label{fig:heating_fig3}
\end{figure}

\begin{figure}[hptb]
\centering
\includegraphics[angle=0,width=8cm,keepaspectratio]{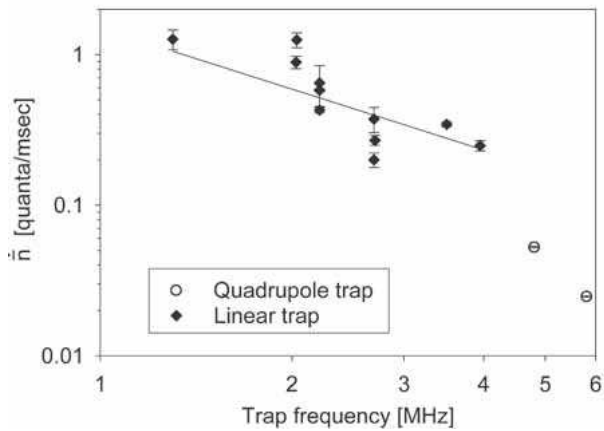}
\caption{Observed heating rates $\dot{\bar{n}}$ vs trap frequency in the $^{111}$Cd$^+$ ion system. The solid line is a fit to the heating rates in the linear trap where the rf electrode voltage $V_{0}$ is constant. The fit to the data in this trap shows a trap frequency scaling of $\omega_{x}^{-2.4}$, yielding a frequency dependence for the spectral density of electric field noise of $S_{E}(\omega)\propto\omega^{-1.4}$. The error bars represent the statistical noise in the linear regression from which a heating rate is inferred (as shown in Fig. \ref{fig:heating_fig3}). The heating measurements were taken over a period of five months in the quadrupole trap and two months in the linear trap. } 
\label{fig:NdotFreqUofM}
\end{figure}

Uncontrolled interactions with the environment can thermally drive the ion to higher motional energies. In order to measure the heating rates, a time delay with no laser interaction is introduced between the ground state cooling and the measurement of $\bar{n}$. These measurements of $\bar{n}$ are repeated with increasing time delay until a heating rate can be extracted. An example of data from the quadrupole trap ($\omega_{x}/2\pi = 5.8$ MHz) is shown in Fig. \ref{fig:heating_fig3}, where a linear fit of the data ($\bar{n}$ vs time delay) yields a heating rate of $\dot{\bar{n}}$= 0.0248(3) quanta/msec. Figure \ref{fig:NdotFreqUofM} displays a series of measured heating rates in the two traps as a function of the trap frequency $\omega_{x}$. In the quadrupole trap, $\omega_{x}$ is varied by changing the rf voltage V$_{0}$ and/or the static voltage U$_{0}$, whereas in the linear trap, $\omega_{x}$ is varied by adjusting only the static potentials on the electrodes.

The dependence of the heating rate on various trap parameters can be modelled by considering a uniform fluctuating electric field $E(t)$ that couples to the ion's motion through its charge. The resulting expression for the heating rate is given by $\dot{\bar{n}}$=$e^{2}S_{E}(\omega_{x})/4m\hbar\omega_{x}$, where $S_{E}(\omega)=2\int^{\infty}_{-\infty}\left\langle E(t)E(t+ \tau)\right\rangle e^{i \omega \tau}d\tau$ is the spectral density of electric field noise in units of (V/m)$^{2}$Hz$^{-1}$ \cite{savard:1997}. It is widely believed that the source of ion heating is noisy ``patch" potentials on the trap electrodes \cite{turchette:2000}, possibly from fluctuating microscopic crystal domains or adsorbed materials on the electrode surface. A simple model of this process suggests an electric field noise $S_{E}(\omega)$ that scales roughly as $1/d^{4}$, where $d$ is the distance between the ion and the closest trap electrode \cite{turchette:2000}. Other potential sources of electric field noise, including thermal (blackbody or Johnson) noise are expected to scale as $S_{E}(\omega)$ $\propto$ $1/d^{2}$ \cite{turchette:2000,lamoreaux:1997,wilkens:1999}.

\begin{figure}[hbpt]
\centering
\includegraphics[angle=0,width=8.6cm,keepaspectratio]{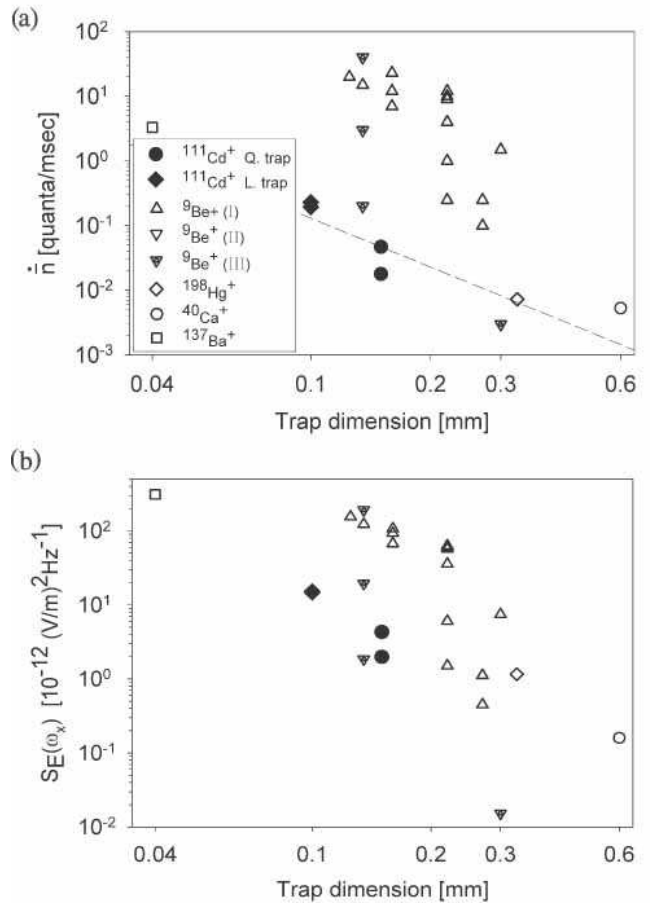}
\caption{(a) Observed heating rate $\dot{\bar{n}}$ vs distance to nearest electrode $d$ in several trapped ion systems with trap frequencies between 2.9 MHz and 6.2 MHz. In addition to the $^{111}$Cd$^{+}$ measurements reported here, previous measurements in $^{9}$Be$^{+}$(I,III) \cite{turchette:2000}, $^{9}$Be$^{+}$(II) \cite{rowe:2001},$^{198}$Hg$^{+}$ \cite{diedrich:1989},$^{40}$Ca$^{+}$ \cite{roos:1999} and $^{137}$Ba$^{+}$ \cite{devoe:2002} are also indicated for reference. Data set $^{9}$Be$^{+}$ (III) represents heating measurements in a particular trap that exhibited anomalously low heating \cite{turchette:2000}. The dashed line is a guide to the eye for the $1/d^{4}$ scaling of the heating rate predicted by a model of microscopic potential fluctuations \cite{turchette:2000}. (b) Spectral density of electric field noise $S_{E}(\omega_{x})$ at the ion inferred from the heating rate in (a) also vs distance to nearest electrode $d$.     
 }
\label{fig:heatingfig6}
\end{figure}

Figure \ref{fig:heatingfig6}(a) displays the heating measurements $\dot{\bar{n}}$ as a function of the distance to the nearest trap electrode $d$ for this experiment ($^{111}$Cd$^{+}$), as well as previously published measurements in $^{198}$Hg$^{+}$ \cite{diedrich:1989}, $^{9}$Be$^{+}$ \cite{monroe1:1995,turchette:2000,rowe:2001}, $^{40}$Ca$^{+}$ \cite{roos:1999}, and $^{137}$Ba$^{+}$ \cite{devoe:2002}; the data is restricted to trap frequencies between 2.9 MHz and 6.2 MHz. Figure \ref{fig:heatingfig6}(b) displays the same data expressed in terms of $S_{E}(\omega_{x})$, or the inferred spectral density of electric field noise present in the various experiments. This provides a useful comparison between systems, since $S_{E}(\omega)$ should not depend on the trapped-ion species. The reported anomalous heating in all trapped ion systems is governed by a process other than thermal noise. In the quadrupole trap, the expected contribution of thermal heating to $S_{E}(\omega)$ is several orders of magnitude smaller than what is measured, while in the linear trap, the expected thermal heating is approximately 20 times smaller than observed (the larger contribution to the thermal noise in the linear trap stems from 1 k$\Omega$ resistors attached to the end-caps for filtering purposes \cite{rowe:2001}). From Fig. \ref{fig:heatingfig6}, we conclude that the heating rates measured in trapped $^{111}$Cd$^{+}$ ions are significantly lower than that measured in other traps of similar dimension and trap frequency. While the relatively large mass of $^{111}$Cd$^{+}$ is partly responsible for this improvement, the inferred electric field noise in the traps reported here still compares favorably to the values obtained in other systems, even though no effort was made to shield the electrodes from the atomic source. 

In summary, we have cooled a single $^{111}$Cd$^{+}$ ion to its ground state of motion and measured heating with different ion trap geometries, materials and strengths. The observed low motional heating, presumably in the presence of Cd coated electrodes, suggests that trapped $^{111}$Cd$^{+}$ ions may be well-suited for experiments involving motional control of trapped ions, including applications in quantum information science.


\end{document}